\documentclass[twocolumn]{aastex701}

\usepackage{xcolor}
\newcommand{\addcolor}{black}

\newcommand{\add}[1]{\textcolor{\addcolor}{#1}}
\newcommand{\remove}[1]{}

\newcommand{\skindepth}{c / \omega_p}
\newcommand{\pparam}{\mathcal{P}}
\newcommand{\pflux}{\mathcal{F}}

\newcommand{\dabs}{\Delta_\text{abs}}

\newcommand{\Ai}[1]{\, \text{Ai}\!\left({#1}\right)}
\newcommand{\Bi}[1]{\, \text{Bi}\!\left({#1}\right)}

\newcommand{\Aip}[1]{\, \text{Ai}'\!\left({#1}\right)}
\newcommand{\Bip}[1]{\, \text{Bi}'\!\left({#1}\right)}

\usepackage{amsmath}

\begin{document}

\title{Transition to Petschek Reconnection in Subrelativistic Pair Plasmas: Implications for Particle Acceleration}


\author[0009-0005-5560-2003]{Adam Robbins}
\affiliation{Department of Astrophysical Sciences, Princeton University, 4 Ivy Lane, Princeton, NJ 08540}
\affiliation{Princeton Plasma Physics Laboratory (PPPL), Princeton, NJ 08540, USA}
\email{arobbins@pppl.gov}

\author[0000-0001-9179-9054]{Anatoly Spitkovsky}
\email{anatoly@princeton.edu}
\affiliation{Department of Astrophysical Sciences, Princeton University, 4 Ivy Lane, Princeton, NJ 08540}

\begin{abstract}

While relativistic magnetic reconnection in pair plasmas has emerged in recent years as a candidate for the origin of radiation from extreme astrophysical environments, the corresponding subrelativistic pair plasma regime has remained less explored, leaving open the question of how relativistic physics affects reconnection. In this paper, we investigate the differences between these regimes by contrasting 2D particle-in-cell simulations of reconnection in pair plasmas with relativistic magnetization ($\sigma \gg 1$) and subrelativistic magnetization ($\sigma < 1$). By utilizing unprecedentedly large domain sizes and outflow boundary conditions, we demonstrate that lowering the magnetization results in a change in the reconnection geometry from a plasmoid chain to a Petschek geometry, where laminar exhausts bounded by slow-mode shocks emanate from a single diffusion region. We attribute this change to the reduced plasmoid production rate in the low-$\sigma$ case: when the secondary tearing rate is sufficiently low, plasmoids are too few in number to prevent the system from relaxing into a stable Petschek configuration. This geometric change also affects particle energization: we show that while high-$\sigma$ plasmoid chains generate power-law energy spectra, low-$\sigma$ Petschek exhausts merely heat incoming plasma and yield negligible nonthermal acceleration. These results have implications for predicting the global current sheet geometry and the resulting energy spectrum in a variety of systems.

\end{abstract}

\section{Introduction} \label{sec:intro}

Reconnection is a ubiquitous plasma process that rapidly converts magnetic energy into particle energy via rearrangement of the magnetic field topology. Originally posited to explain solar flares, reconnection now plays a key role in the theories of a multitude of astrophysical systems, from the heliosphere to compact objects. Observations and simulations suggest that reconnection can energize particles far beyond thermal energies, generating hard power-law energy spectra \citep{sironi2014, guo2014}. The conditions under which this particle acceleration occurs, the precise mechanisms responsible, and the shape of the resultant spectra have been intensively studied in recent years with the aid of particle-in-cell (PIC) simulations \citep{li2021, guo2024, sironi2025}. These simulations have enabled the study of kinetic reconnecting plasmas on increasingly large spatial and temporal scales.  

One of the key determinants of particle acceleration is the current sheet geometry: each particle acceleration mechanism requires certain configurations of electric and magnetic fields to operate. There are two broad categories 
of reconnecting current sheet geometries: single X-point and multiple X-point\footnote{
also referred to as X-lines, the three-dimensional analogues of X-points.} \citep{ji2011}. In the former, there is just one diffusion region where the field lines break and reconnect, while most of the reconnecting plasma proceeds directly from the inflow to the outflow exhaust. In the latter, the current sheet fragments due to tearing or plasmoid instability, resulting in a dynamic chain of magnetic islands with X-points in between \citep{nuno2007}.\footnote{In this paper we shall use X-point and diffusion region interchangeably to refer to areas where non-ideal electric fields dissipate inflowing oppositely-aligned magnetic flux.}
\remove{Whether reconnection proceeds with a single X-point or with multiple X-points is critical to understanding particle acceleration dynamics.} Recent work has established the importance of direct electric field acceleration at X-points in relativistic plasmas as an injection mechanism for subsequent acceleration \citep{sironi2022, totorica2023, gupta2025}, a pathway which is more accessible in systems with more X-points. Furthermore, plasmoid-based Fermi acceleration mechanisms at later stages of energization \citep{drake2006, guo2019, hakobyan2021} are also primarily operative in the multiple X-point regime. In contrast, pickup acceleration occurs mainly in the single X-point regime for particles that do not pass through the X-point \citep{drake2009}. \remove{How these mechanisms interact to produce the observed total spectra is a key research question.}

Given the strong dependence of particle acceleration on the current sheet geometry, it would be useful to have a framework to predict the transition from single X-point reconnection to multiple X-point reconnection. In the MHD model, this is accomplished by the Lundquist number $S \equiv L v_A / (\eta c / 4\pi)$, where $L$ is the half-length of the current sheet, $v_A$ is the Alfv\'en speed, and $\eta$ is the resistivity, assumed to be constant. When $S > S_c \sim 10^4$, the reconnecting current sheet fragments, resulting in fast plasmoid-mediated multiple X-point reconnection \citep{ji2011}.\remove{Of course, this fluid model does not always carry over to collisionless plasmas.} Although recent strides have been made in understanding the tearing properties of collisionless current sheets \citep{hoshino2020, schoeffler2024, akutagawa2025}, there is no comparable metric to \textit{a priori} predict the global reconnection geometry in collisionless systems, suggesting the need for kinetic studies. 

These questions also tie in closely to debates concerning the origin of fast reconnection. \citet{petschek1964} proposed that the slow Sweet-Parker reconnection \citep{parker1957} could be made fast by flanking a small diffusion region with four slow-mode shocks responsible for the  majority of energy dissipation. His model has proven controversial since, in the absence of additional physics to localize the diffusion region, the Petschek solution is unstable and devolves to that of Sweet and Parker in MHD simulations \citep{uzdensky2000, forbes2013}. On the other hand, slow-mode shocks have been observed in the Earth's magnetotail for decades \citep[e.g.,][]{feldman1987, saito1995, walia2024}, providing observational evidence in favor of Petschek's theory. Large-domain kinetic simulations would thus be well suited to address whether Petschek-type outflows can form, and, if so, under what conditions. 

Subrelativistic pair plasmas for which $\sigma < 1$ are ideal systems for studying the physics of the transition from multiple to single X-point reconnection. The magnetization $\sigma$ is defined as 

\begin{equation}
    \sigma = \frac{B_0^2}{4 \pi n_0 mc^2},
\end{equation} where $B_0$ and $n_0$ are the magnetic field and number density of the upstream plasma, respectively. The magnetization represents the characteristic Lorentz factor that particles would gain if all the magnetic energy were dissipated into cold particles. Thus, $\sigma \gg 1$ indicates relativistic reconnection and $\sigma < 1$ the subrelativistic regime. High-$\sigma$ systems are characterized by rapid plasmoid production, growth, and mergers \citep{sironi2016}. Since the rate of plasmoid formation increases with increasing $\sigma$, by lowering $\sigma$ below 1 we can address systems with lower tearing rates than those typically studied for pair plasmas. The result, as we shall demonstrate, is that while high-$\sigma$ simulations always yield multiple X-point plasmoid chains, those with low-$\sigma$ instead have just a single X-point. By comparing otherwise identical simulations that differ only in $\sigma$, we can understand the bifurcation of the system into single or multiple X-point regimes and study how this bifurcation affects the resultant particle acceleration. 

\remove{The use of electron-positron plasmas for this purpose is advantageous as it both enables large-scale simulations and facilitates the comparison of our results with the existing literature on high-$\sigma$ pair reconnection.  However, we do expect our results to carry over to electron-ion plasmas, as we shall discuss later.}

In this investigation, we focus on two magnetization values: $\sigma = 0.3$ and $\sigma = 10$, which have (upstream) Alfvén speeds $v_A = c \sqrt{\sigma / (\sigma + 1)}$ of $\approx 0.48c$ and $\approx 0.95c$, respectively. The $\sigma = 10$ case is well within the relativistic reconnection regime and, as we shall see, the resulting reconnection rate, current sheet structure, and particle energy spectra are consistent with the existing literature \citep[e.g.,][]{sironi2016, guo2024}. This acts as a control with which we may contrast the $\sigma = 0.3$ case. Although we focus on only these two cases, we expect that the key results for $\sigma = 10$ also apply to $\sigma > 10$, and the results of $\sigma = 0.3$ apply to $\sigma < 0.3$, with the range $0.3 < \sigma < 10$ presenting an intermediate regime.

Our main finding is that the global structure of the current sheet is determined by the tug-of-war between the formation of wide-angle exhausts, which are naturally established to satisfy mass conservation, and plasmoid formation, which tends to perturb or disrupt those exhausts. If the number of plasmoids streaming forth from the diffusion regions is sufficiently large to prevent the exhausts from forming, the entire domain is thereby made available for multiple X-point reconnection and its associated mechanisms of particle acceleration (namely, direct $E$-field, Fermi, and betatron). However, if the plasmoid activity is insufficient to prevent the exhausts from being established, one is left with just a single X-point. The aforementioned canonical particle acceleration mechanisms are thereby throttled, yielding just a thermal Maxwellian distribution in particle energy. 
 
This paper is organized as follows. In Section \ref{sec:setup}, we discuss the setup of our simulations. In Section \ref{sec:current_sheet}, we discuss how the change in magnetization affects the global reconnection geometry. In Section \ref{sec:particle_acceleration}, we discuss how those changes in the current sheet structure alter the particle acceleration pathways and resultant energy spectrum. In Section \ref{sec:conclusion}, we recapitulate our basic conclusions and discuss broader implications. 

\section{Simulation Setup} \label{sec:setup}

We perform 2D simulations of reconnection using the electromagnetic PIC code TRISTAN-V2 \citep{tristan}. For all simulations, the mass ratio $m_i / m_e = 1$, and there are 16 total particles per cell. The speed of light is set to $c = 0.45$ cells per time step to satisfy the CFL condition. We vary the magnetization, skin depth, and domain size as described in Table \ref{tab:sims}. 

\begin{table}
    \centering
    \begin{tabular}{|l|r|r|r|r|}
        \hline
        Run & $\sigma$ & $\skindepth \; [\Delta x]$ & $L_x \; [\skindepth]$ & $L_y \; [\skindepth]$ \\
        \hline
         I & 10 & 50 & 120 & 240 \\
         \hline
         II & 0.3 & 50 & 120 & 240 \\
         \hline
         III & 10 & 25 & 240 & 480 \\
         \hline
         IV & 0.3 & 25 & 240 & 480\\
         \hline
         V & 0.3 & 8 & 750 & 6000 \\
         \hline
    \end{tabular}
    \caption{Simulation parameters}
    \label{tab:sims}
\end{table}

\par 
All simulations begin with a Harris sheet equilibrium of the form

\begin{align}
    &\bold{B} = B_0 \tanh{(x / \Delta)} \; \bold{\hat{y}}, \\
    &\add{{n = n_0 \left(1 + 3 \: \text{sech}^2{(x / \Delta)} \right)}},
\end{align} where $\Delta$ is set to 100 cells. \remove{Note that since no guide field is included, this is anti-parallel reconnection. The overdensity in the current sheet is initially $3 n_0$.} Reconnection is triggered by density depletion in the center of the current sheet, sending reconnection fronts outward in both directions. In the $x$-direction, fresh particles and flux are constantly supplied to replenish the reconnected plasma. These inflows are initialized cold, i.e., with $\theta \equiv T / mc^2 = 0$. In the $y$-direction, we use outflow boundary conditions which permit the reconnected plasma to be advected out of the system \add{(for details, see Appendix \ref{app:bcs})}. These outflow boundary conditions crucially allow the system to reach steady-state equilibrium, a process which generally takes several Alfv\'en crossing times. This lengthy equilibration period is used to eliminate any sensitivity of our results to the details of the initial current sheet and to ensure that we are studying long-term behavior rather than initial transients. 
\par 
We run simulations with a variety of spatial resolutions and domain sizes, summarized in Table \ref{tab:sims}. The moderate-size and moderate-resolution Runs III and IV are used to illustrate the essential comparison between high- and low-$\sigma$. The small, highly resolved Runs I and II are used when considering effects at the scale of the diffusion region, such as X-point particle acceleration and the tearing rate within the diffusion region. Run V, covering the largest extent, is used to verify that the patterns observed in smaller subrelativistic simulations (Runs II and IV)  carry over to very large system sizes. Ultimately, all simulations for each value of $\sigma$ show the same basic current sheet structure and particle spectrum---we do not observe any differences from varying the skin depth, implying that the higher resolutions used in Runs I-IV were unnecessary to capture the fundamental kinetic physics. 

\section{Current Sheet Geometry} \label{sec:current_sheet}

\subsection{Petschek-like Exhausts}

\begin{figure}
    \centering
    \includegraphics[width=1.0\linewidth]{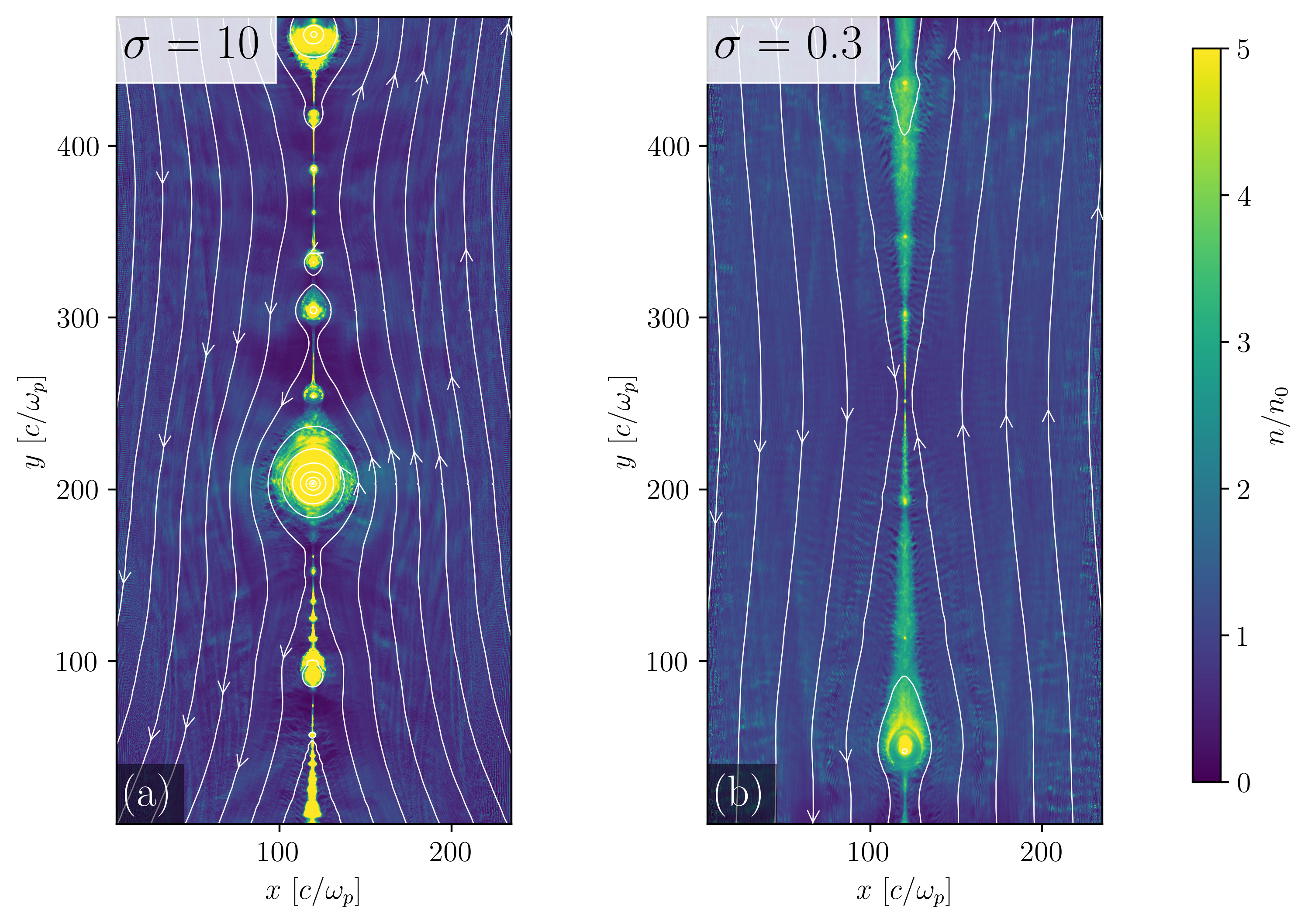}
    \caption{Density profiles for Runs III and IV, respectively, once steady state is achieved. The density scale is cut off at five times the inflow density. }
    \label{fig:summary_density}
\end{figure}

Reducing the magnetization from $\sigma = 10$ to $\sigma = 0.3$ results in a stark change in the structure of the current sheet, as can be seen in Figure~\ref{fig:summary_density}. In the $\sigma = 10$ case, as expected, the tearing instability results in copious plasmoid formation, frequent plasmoid mergers, and rapid plasmoid growth. The result is a self-similar current sheet made up of plasmoids of various sizes, as described in detail in \citet{sironi2016}. This is a clear case of multiple X-point reconnection wherein field lines reconnect at many locations along the current sheet. However, a primary X-point can be identified at $y \approx 360 \; \skindepth$ during this timestep. 

\par 
The situation for $\sigma = 0.3$ is different. The current sheet near the X-point at $y = 240 \; \skindepth$ is also tearing-unstable, resulting in the regular production of plasmoids. However, these plasmoids seldom merge and only a handful are present in the domain at any given time. The large-scale current sheet is a Petschek-like configuration: there is a small diffusion region spanning several skin depths in the center with exhausts at either end. By exhausts we refer to the moderate-density, weak-magnetic-field regions in the downstream, i.e., as distinguished from the plasmoids and the X-point(s). 

\begin{figure}
    \centering
    \includegraphics[width=0.8\linewidth]{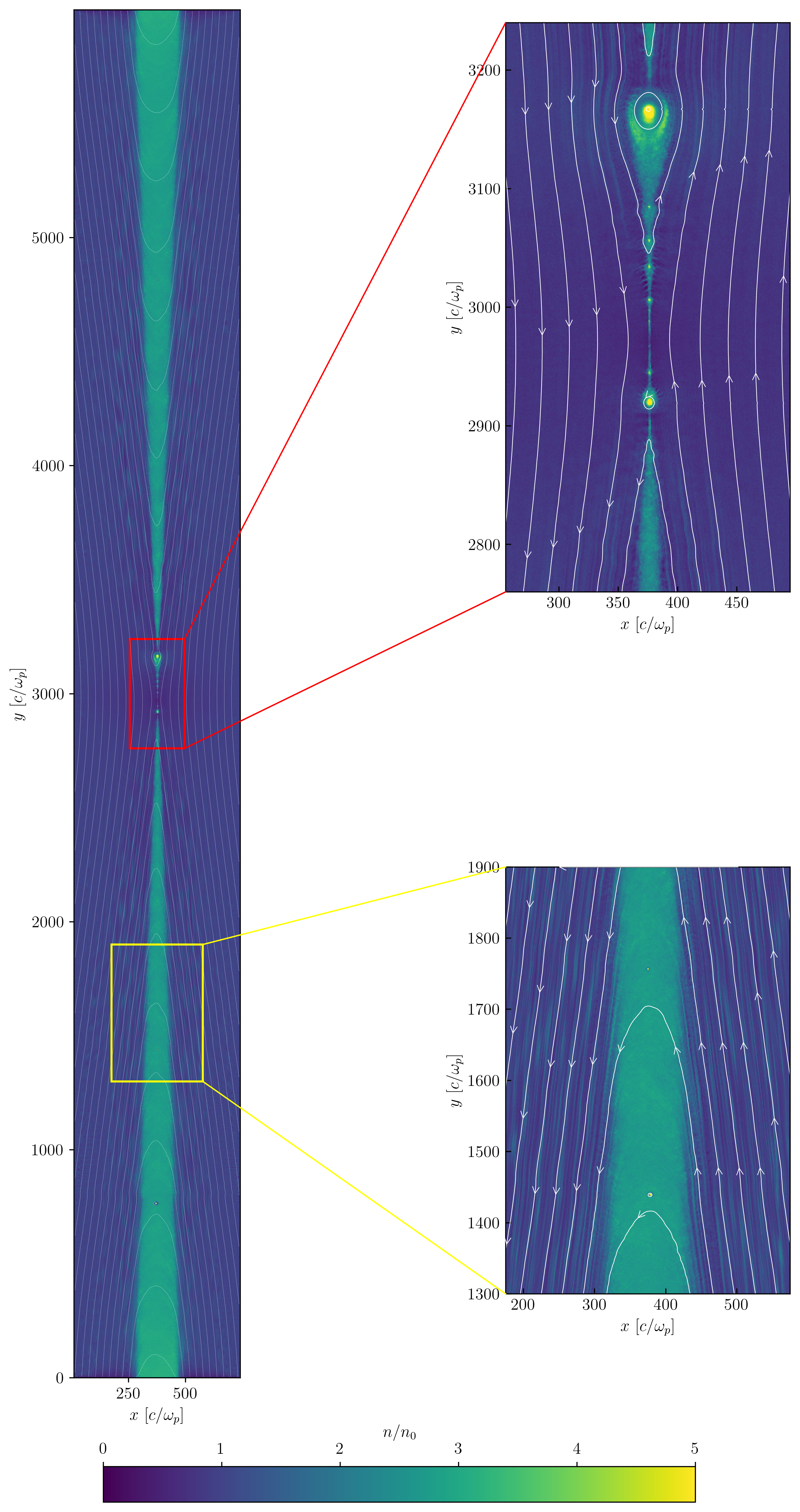} 
    \caption{Density profile for the $\sigma = 0.3$ Run V spanning $6000 \; c/\omega_p$ in the $y$-direction once it has reached steady state. Note that while plasmoids near the X-point have comparable widths to  the width of the exhaust, by the time the plasmoids move farther away from the X-point their width becomes much smaller than that of the exhaust, and they become comparatively insignificant. Two plasmoids can be seen at $y \approx 1440$ and $1760 \; \skindepth$. An animated \href{https://youtube.com/playlist?list=PL-1sbkerHVc_Cxk0X2DX6jZhpSMV1ktlM&si=hcuGYkDMQXw76oGL}{version} of this figure \add{shows the steady-state evolution over several Alfv\'en-crossing times.}}
    \label{fig:bigrun}
\end{figure}

The exhausts are even more pronounced in Run V that has a domain size ten times longer in the $y$-direction, shown in Figure~\ref{fig:bigrun}. Here we see that when $L_y \gg c/\omega_p$, the plasmoids become nearly insignificant compared to the large laminar exhausts that form. While plasmoids do perturb these exhausts \add{(as noted in \citet{innocenti2015})}, they are too few in number to disrupt them: once the laminar exhausts stabilize, they persist for as long as we are able to simulate. Reconnection is still fast with the reconnection rate $R \equiv v_\text{in} / v_A \approx 0.10$, the same as for Run IV, and only somewhat lower than $R \approx 0.15$ for $\sigma = 10$ (Run III). 

\remove{The relative importance of plasmoids can be quantified by the plasmoid energy fraction, which we define as the quotient of the particle kinetic energy within plasmoids to the total particle kinetic energy within the overall current sheet. A region is considered part of a plasmoid if it is fully enclosed by a closed field line. The plot of the plasmoid energy fraction for each run is shown in Figure~\ref{fig:pfrac}. Even in small runs that capture dynamics near the X-point, there is a clear difference between $\sigma = 10$ runs where much more than 90\% of the total energy is within plasmoids compared to less than 25\% for $\sigma = 0.3$. For $\sigma = 0.3$, as the size of the domain increases, this fraction decreases while remaining roughly constant for the case $\sigma = 10$.}

\begin{figure}
    \centering
    \includegraphics[width=0.75\linewidth]{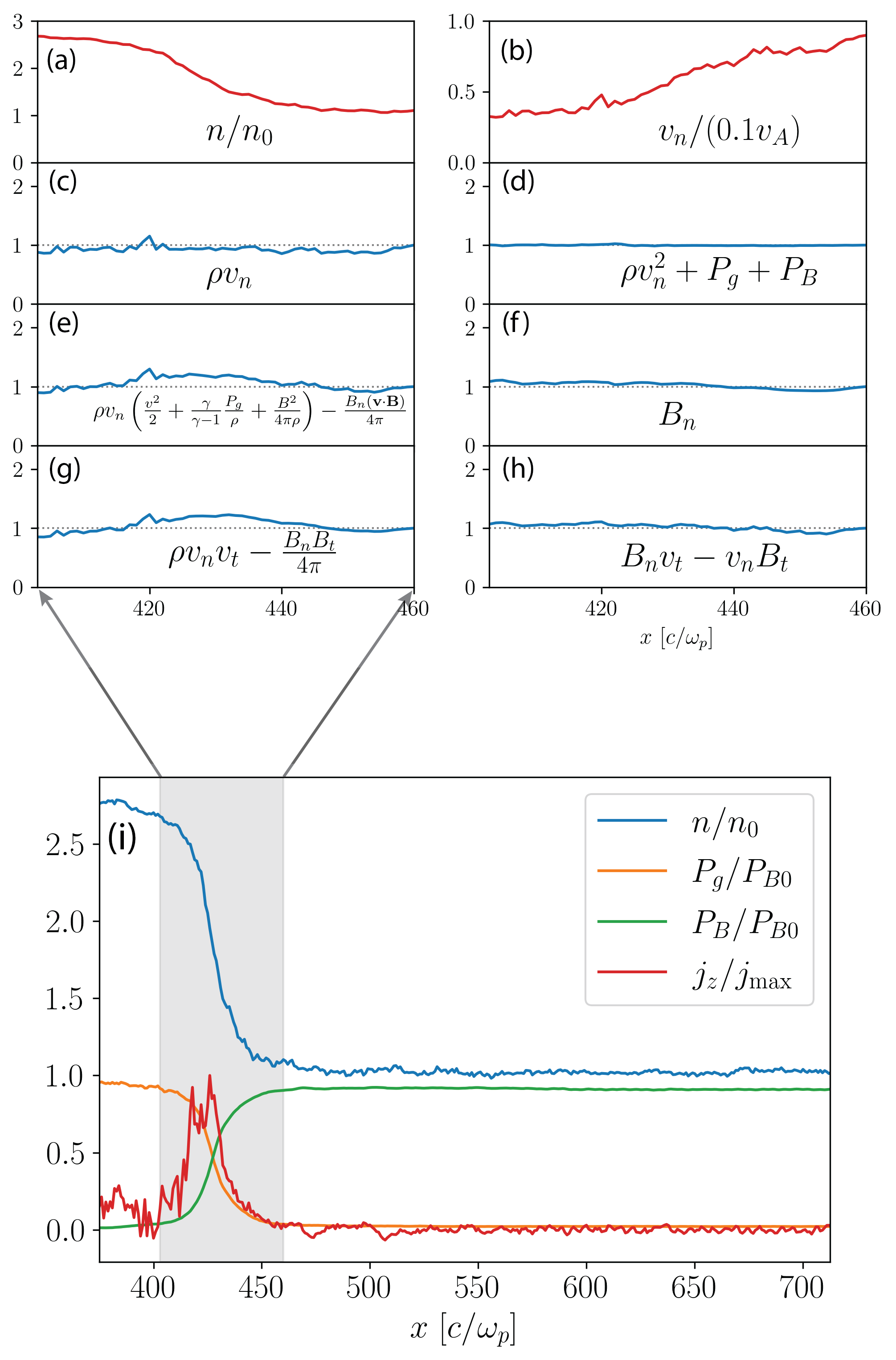}
    \caption{Plots (a) - (h) show selected quantities for a cut along $y = 1600 \; c/\omega_p$ in Run V. Plots (a) and (b) show the normalized density and normal bulk velocity, respectively. Plots (c) - (h) show the Rankine-Hugoniot jump quantities, normalized to their upstream values.  Plot (i) shows general quantities plotted along that same $y$-value; the transition region depicted in plots (a) - (h) is shaded in gray. The upstream magnetic pressure $P_{B0} \equiv B_0^2 / 8 \pi$.}
    \label{fig:combined_cuts}
\end{figure}

Thus, for low magnetization, the vast majority of inflowing particles avoid plasmoids altogether and simply proceed directly from the cold inflow to the hot exhaust. At this interface slow-mode shocks form, as can be confirmed via comparison with the six Rankine-Hugoniot jump conditions representing continuity in mass, normal momentum, energy, normal magnetic field, transverse
momentum, and tangential electric field, respectively \citep{walia2024}:

\begin{equation}
    [\rho v_n] = 0,
\end{equation}

\begin{equation}
    \left[\rho v_n^2 + P_g + \frac{B^2}{8 \pi}\right] = 0,
\end{equation}

\begin{equation}
    \left[\rho v_n \left(\frac{v^2}{2} + \frac{\gamma}{\gamma - 1} \frac{P_g}{\rho} + \frac{B^2}{4 \pi \rho}\right) - \frac{B_n (\mathbf{v} \cdot \mathbf{B})}{4 \pi}\right] = 0,
\end{equation}

\begin{equation}
    [B_n] = 0,
\end{equation}

\begin{equation}
    \left[\rho v_n {\mathbf{v}}_t - \frac{B_n \mathbf{B}_t}{4 \pi}\right] = 0,
\end{equation}

\begin{equation}
    [B_n {\mathbf{v}}_t - v_n \mathbf{B}_t] = 0.
\end{equation} Here $\rho$ represents the mass density, $\mathbf{v}$ the flow velocity, $P_g$ the gas pressure, and $\gamma = 5 / 3$ the adiabatic index. The components normal and tangential to the shock are denoted by $n$ and $t$, respectively. Figure~\ref{fig:combined_cuts} shows these quantities along a horizontal cut across the exhaust boundary, far from the X-point. As one can see, these jump conditions are well satisfied in our simulations, in general agreement with the Petschek model \add{(see \citet{walia2022} for a similar analysis of a hybrid simulation)}. Note, however, that, contrary to \citet{petschek1964}, these slow-mode shocks are not also switch-off shocks, consistent with earlier work \citep{liu2012}. This is because some current is spread throughout the exhaust rather than being localized only at the shock front (Fig~\ref{fig:combined_cuts}), resulting in the field lines gradually curving in the exhaust rather than bending sharply (Figure~\ref{fig:bigrun}). Across these slow-mode shocks, the magnetic pressure is converted into particle pressure and the density rises to $n / n_0 \approx 3$. Mass conservation allows us to determine the approximate outflow angle $\theta$, as a function of this ratio and the reconnection rate $R \approx 0.1$:

\begin{equation}
    \tan{{\frac{\theta}{2}}} = R \frac{n_0}{n}. \label{eq:masscon}
\end{equation} This results in an outflow angle of $\theta \approx 4^\circ$, consistent with the simulation (Figure~\ref{fig:bigrun}). \add{While the relativistic case (Fig. \ref{fig:summary_density}a) does not have this sort of persistent exhaust, one can note smaller ``minijets"  trailing the plasmoids as they advect out \citep{macias2020}. The minijets display a drop in tangential B field across their boundary similar to a slow-shock transition.}

\subsection{Single to Multiple X-point Transition}

In sharp contrast to this Petschek configuration, all of our $\sigma = 10$ simulations (runs I and III), as well as those in larger 2D domains \citep[e.g.,][]{sironi2016}, show large plasmoid chains with no significant exhausts. This sharp discrepancy is caused by the interplay between two competing processes operative in the current sheets. The first is the natural tendency for a reconnecting system to form wide-angle exhausts to ensure flow continuity (as in Equation (\ref{eq:masscon})). In these exhausts, the incoming flow is heated and redirected to the outflow direction. The second is the fracturing of the thin current sheets in the diffusion regions into plasmoids as part of the secondary plasmoid tearing instability \citep{uzdensky2010, sironi2016}. These plasmoids, if sufficiently large and numerous, can disrupt the exhausts by physically displacing them. Conversely, the exhausts can suppress plasmoid formation because they widen the current sheet and thereby prevent further tearing.

\begin{figure}
    \centering
    \includegraphics[width=1\linewidth]{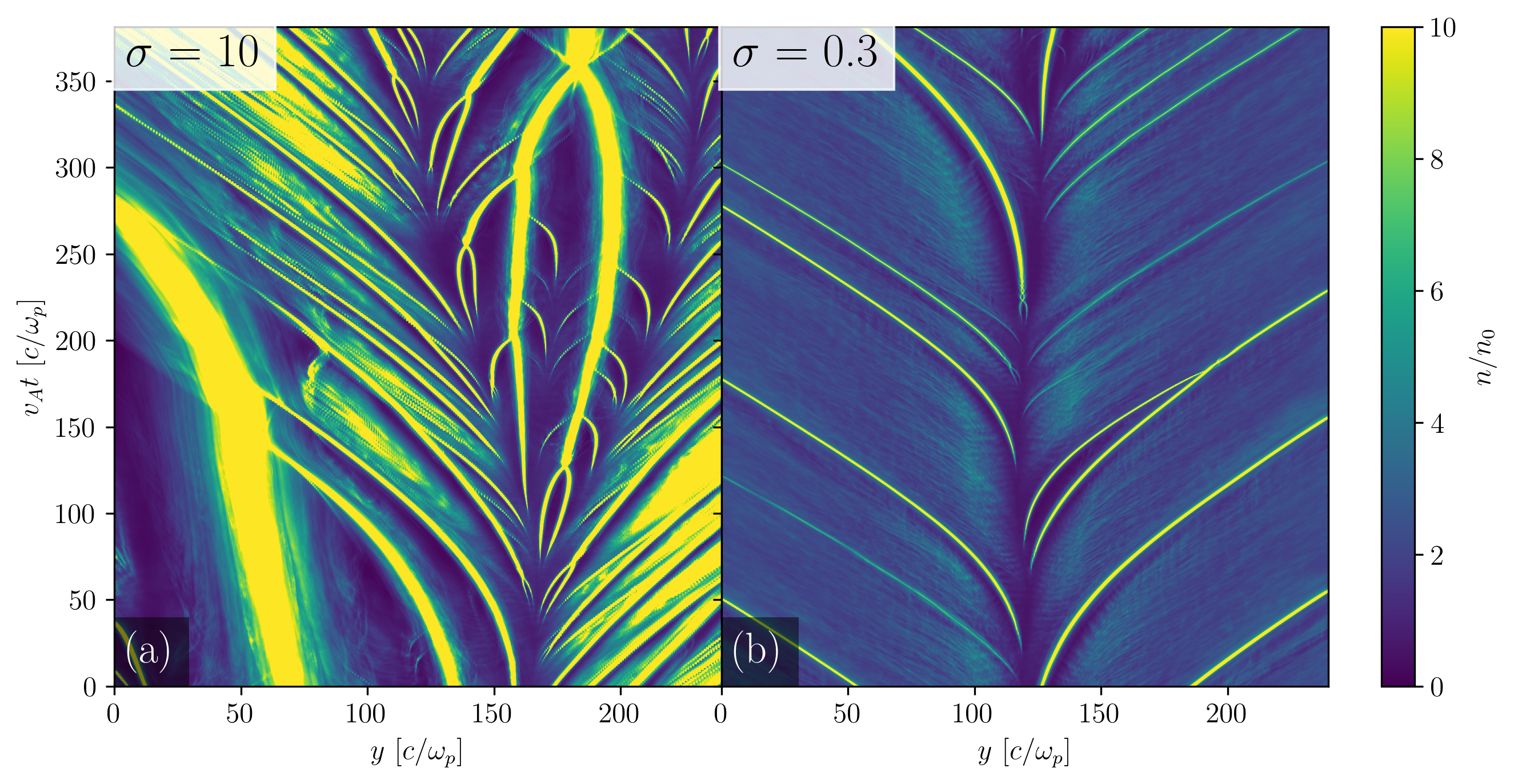}
    \caption{Spacetime plots of the midline ($x = L_x / 2 = 60 \; \skindepth$) for Runs I and II respectively after they have reached equilibrium. The areas with the lowest density (the darkest blue) are the diffusion regions and the high-density streaks of yellow are the plasmoids---thus these plots show the life journey of plasmoids as they are born and leave the domain. As one can see in the $\sigma = 0.3$ case, plasmoid production, while regular, is sufficiently infrequent to allow most plasmoids to escape the \remove{region near }the X-point at the center without ever merging with another plasmoid (though one prominent collision can be seen near (190, 180)). The case in Run I is quite different - here we see rapid plasmoid production and, crucially, many mergers. This results in large, slow plasmoids which themselves are more susceptible to further mergers. This feedback mechanism fills the domain with large, massive plasmoids. }
    \label{fig:spacetime}
\end{figure}

To see why the $\sigma = 10$ case is plasmoid-dominated and the $\sigma = 0.3$ case is exhaust-dominated, let us consider spacetime diagrams showing the birth, growth, and outflow of plasmoids in the current sheet (Figure~\ref{fig:spacetime}). The key difference between the two regimes is in plasmoid mergers, the process by which two plasmoids collide and form a larger plasmoid \citep[for extensive analysis, see][]{sironi2016}. In the $\sigma = 10$ case (Figure~\ref{fig:spacetime}(a)) this occurs constantly, with each plasmoid undergoing many mergers within its lifetime. In contrast, in the $\sigma = 0.3$ case (Figure~\ref{fig:spacetime}(b)) plasmoids seldom collide---the vast majority of plasmoids are simply born and then advect out of the diffusion region. Without mergers, these plasmoids never become very large---they simply grow modestly and then leave at the Alfv\'en speed (see also Figure~\ref{fig:bigrun}). While the steady stream of numerous large plasmoids in the $\sigma = 10$ case swiftly disrupts any nascent exhausts, the trickle of small plasmoids in the $\sigma = 0.3$ is sufficiently feeble that exhausts can form unimpeded. \remove{Although the plasmoids in the latter case do perturb the exhaust as they pass through (see Figure~\ref{fig:summary_density}b and \citet{innocenti2015}), they are too few in number to permanently disperse it. Furthermore, these effects are self-reinforcing. The plasmoid-dominated current sheet that forms in the high-$\sigma$ case spawns secondary X-points which produce yet more plasmoids. The wide-angle exhausts which form in the low-$\sigma$ case thicken the current sheet away from the X-point, precluding further tearing. The plasmoids passing through the exhausts, immersed in steady flow, neither accrete fresh flux nor merge with each other. Instead, they are simply swept out of the system.}

To quantify this phase transition, we define a dimensionless parameter $\pparam$ which scales as the ratio between the plasmoid width and the inter-plasmoid distance, which we term the \textit{plasmoid parameter}. As $\pparam$ rises, so too does the chance for mergers. Hence for high $\mathcal{P}$ we expect multiple X-point reconnection, and for low $\pparam$ we expect instead single X-point reconnection. If we define $\pflux$ as the plasmoid production rate at the X-point measured in exiting plasmoids per unit time, the mean inter-plasmoid distance will scale as $v_A / \pflux$. The width of a plasmoid upon leaving the diffusion region is roughly that of the current sheet itself, $w_{cs}$. Thus our plasmoid parameter may be expressed as 
\begin{equation}
    \pparam \equiv \frac{\pflux w_{cs}}{v_A}. \label{eq:pparam}
\end{equation} \add{Using the measured values of $\pflux$ and $w_{cs}$, we find $\pparam = 0.10$ for Run I ($\sigma = 10$) and $\pparam = 0.05$ for Run II ($\sigma = 0.3$).} We can thus conclude that $\pparam \gtrsim 0.10$ corresponds to multiple X-point reconnection and $\pparam \lesssim 0.05$ to single X-point, separated by a critical transition parameter $\pparam_c \sim 0.075$. As $\pparam$ increases from $0.05$ to $0.10$ the stream of plasmoids intensifies as more mergers occur until any potential exhaust is rapidly dispersed. Since this expression is not unique to the particular system properties, we expect it to hold in electron-ion plasmas as well where similar current sheet structures are indeed observed \citep[e.g.,][]{rowan2017}.

Lastly, we consider why $\pparam$ rises with magnetization. As $\sigma$ increases, so does the current density within the current sheet. As a result, nearby current elements attract each other more strongly via the Lorentz force, leading to more rapid plasmoid coalescence. We may construct a simple scaling for $\pparam$ as follows. The force between nearby current elements scales as $J^2$, and the relativistic mass density as $\langle \gamma \rangle n$. Thus, the characteristic acceleration rate of these current elements towards each other will scale as 

\begin{align}
    a \propto \frac{J^2}{\langle \gamma \rangle n}.
\end{align} To express this as a function of the magnetization, we make the following assumptions supported by our simulations. Firstly, we assume that the current sheet width $w_{cs}$ and density $n$ are independent of $\sigma$. This implies that $J^2 \propto B_0^2 \propto \sigma$. Secondly, we take $\langle \gamma \rangle \approx 1 + \sigma / 4$, the injection energy identified by \citet{totorica2023}. This results in the acceleration scaling as 

\begin{align}
    a \propto \frac{J^2}{\langle \gamma\rangle n} \propto\frac{\sigma}{1 + \sigma /4}.
\end{align} The inverse time scale associated with this acceleration is $t^{-1} \sim \sqrt{a / w_{cs}}$. This yields our scaling for the plasmoid production rate $\pflux \propto \sqrt{a / w_{cs}} \propto \sqrt{a}$.  Dividing this by $v_A = c \sqrt{\sigma / (\sigma + 1)}$ yields the following scaling for $\pparam$:

\begin{align}
    \pparam \propto \frac{\pflux}{v_A} \propto \sqrt{\frac{1 + \sigma}{1 + \sigma / 4}}.
\end{align} This yields a factor of $1.6$ increase from $\sigma = 0.3$ to $\sigma = 10$, roughly consistent with the data. 

To summarize: when $\sigma$ is reduced from 10 to 0.3, the lower current density results in a lower tearing rate (even when normalized to $v_A$). From this comes fewer plasmoids and fewer plasmoid collisions, resulting in just a trickle of small plasmoids. This relatively modest plasmoid activity is insufficient to stop the formation of exhausts which thereafter dominate the current sheet and result in a single X-point Petschek configuration. 

\section{Particle Acceleration} \label{sec:particle_acceleration}

Our motivation for investigating the transition from the multiple X-point regime to the single X-point regime is to assess the effect of changes in the current sheet geometry on particle acceleration. Now we discuss the particle energy spectrum in both cases and explain why they differ---in particular, how the energization pathways in the $\sigma = 0.3$ case differ from those in the canonical $\sigma = 10$ case. 

\subsection{Total Spectra}

\begin{figure}
    \centering
    \includegraphics[width=1\linewidth]{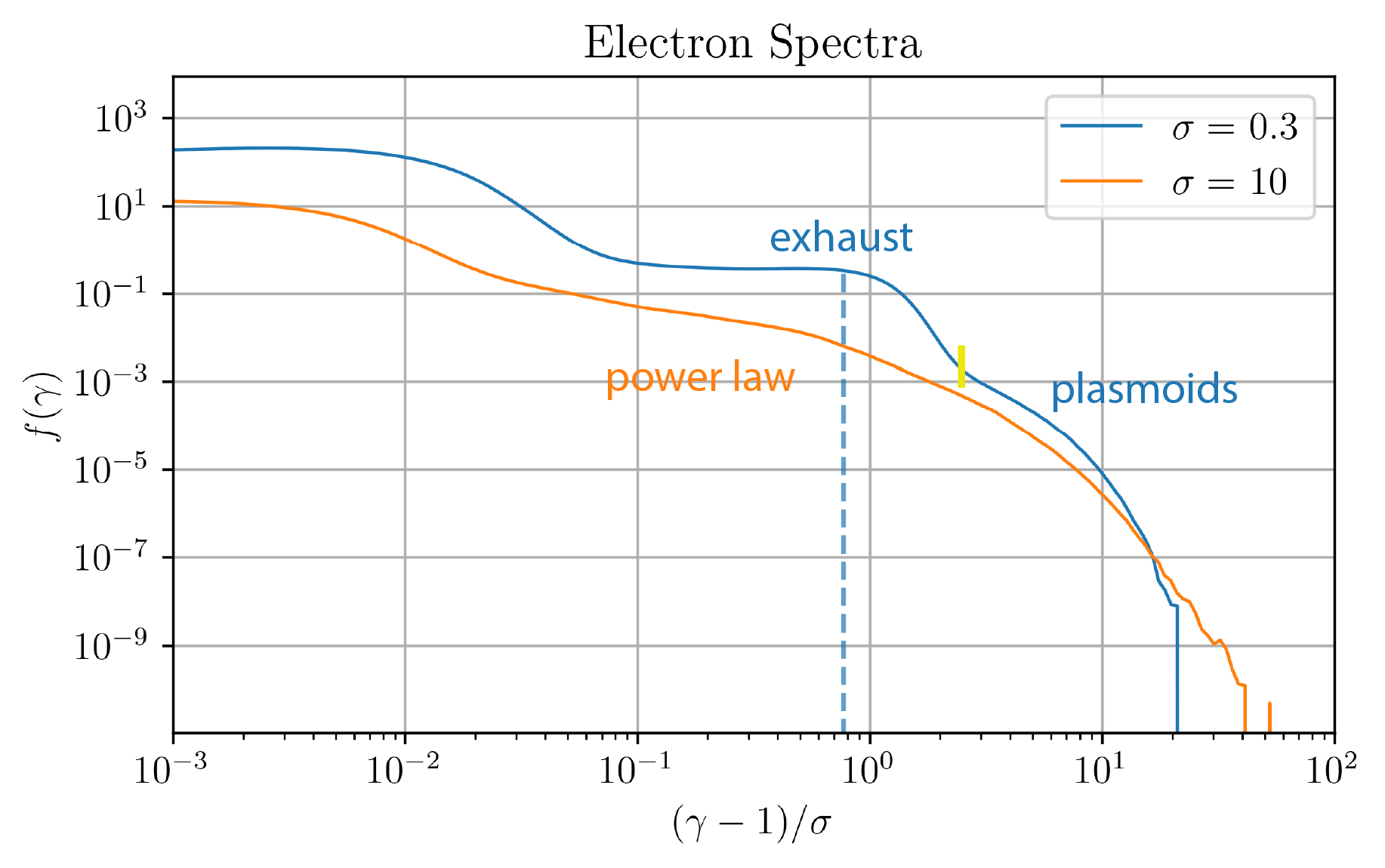}
    \caption{Electron energy spectra for Runs III and IV for the entire domain, normalized by the magnetization. \add{The dashed blue line indicates the expected pickup energy, $m v_A^2$.} The transition between the exhaust component and the plasmoid component is indicated by the \add{yellow line}. Note that the first structures on the far left for each case represent the inflow which has acquired a small thermal temperature due to wave heating. 
    As the domain size ($L_y$) grows for $\sigma = 0.3$, the shape of the plot looks the same, however the magnitude of the plasmoid component in the spectrum decreases relative to that of the exhaust due to the decreasing fraction of the domain occupied by plasmoids.}
    \label{fig:spectra}
\end{figure}

The electron energy spectra---specifically, the electron kinetic energy normalized to the magnetization---are compared in Figure~\ref{fig:spectra}. In the $\sigma = 10$ case \citep[e.g.,][]{sironi2014, guo2014}, the plasmoid-dominated reconnection results in the establishment of a broad, hard power law (that is, at energies greater than those of the cold inflow component seen on the left of Figure~\ref{fig:spectra}). For the $\sigma = 0.3$ case, we see two distinct components in the energized part corresponding to particles within the current sheet. First, on the left, we see a shoulder comprised of a flat spectrum followed by a rapid drop-off. At the end we see a small high-energy tail containing the most energetic particles which branches off from the shoulder at $\gamma \approx 1.7$. These features are caused by distinct particle acceleration mechanisms: the first ``shoulder" component consists of particles that are energized in the exhaust and the second, more energetic, component consists of particles confined in plasmoids.

In relativistic reconnection, particle  acceleration is fundamentally stochastic---particles can transition from direct X-point acceleration to Fermi acceleration within plasmoids and back again. This results in the development of a smooth power-law spectrum. By contrast, in the $\sigma = 0.3$ case, the regular, laminar structure of the reconnection region results in two distinct acceleration pathways which depend on where the particle first encounters the current sheet. The first category of particles follows the field lines into the diffusion region. There they are accelerated by the non-ideal electric fields \citep{bessho2012} before being confined and energized by a compressing plasmoid \citep{hakobyan2021}. The much larger second category of particles, instead of first encountering the current sheet at the X-point, does so at the slow-mode shocks at the boundary of the exhaust outflow. There they are pickup-accelerated and brought up to speed with the Alfv\'enic outflow, resulting in a thermal distribution. It is this latter pathway which ultimately dominates, especially when one considers realistic system sizes where $L_y \gg \skindepth$.

\subsection{X-point Injected Particles} \label{sec:xspec}

Recent work on particle acceleration in  relativistic reconnection has demonstrated that the most energetic particles undergo a two-step acceleration process: injection to relativistic energies via direct acceleration at X-points and then strong energization by Fermi/betatron acceleration powered by ideal electric fields in plasmoids \citep{sironi2022, totorica2023, gupta2025}. Some particles which fall into the x-point undergo an analogous process in $\sigma = 0.3$ reconnection, accelerated by the non-ideal electric field as they undergo Speiser motion \citep{speiser1965}. The result is the establishment of a power-law distribution in the vicinity of the X-point. These particles then are quickly trapped by plasmoids where they ultimately gain most of their energy by Fermi/betatron acceleration as the plasmoid compresses and its magnetic fields intensify. 

\begin{figure}
    \centering
    \includegraphics[width=1.0\linewidth]{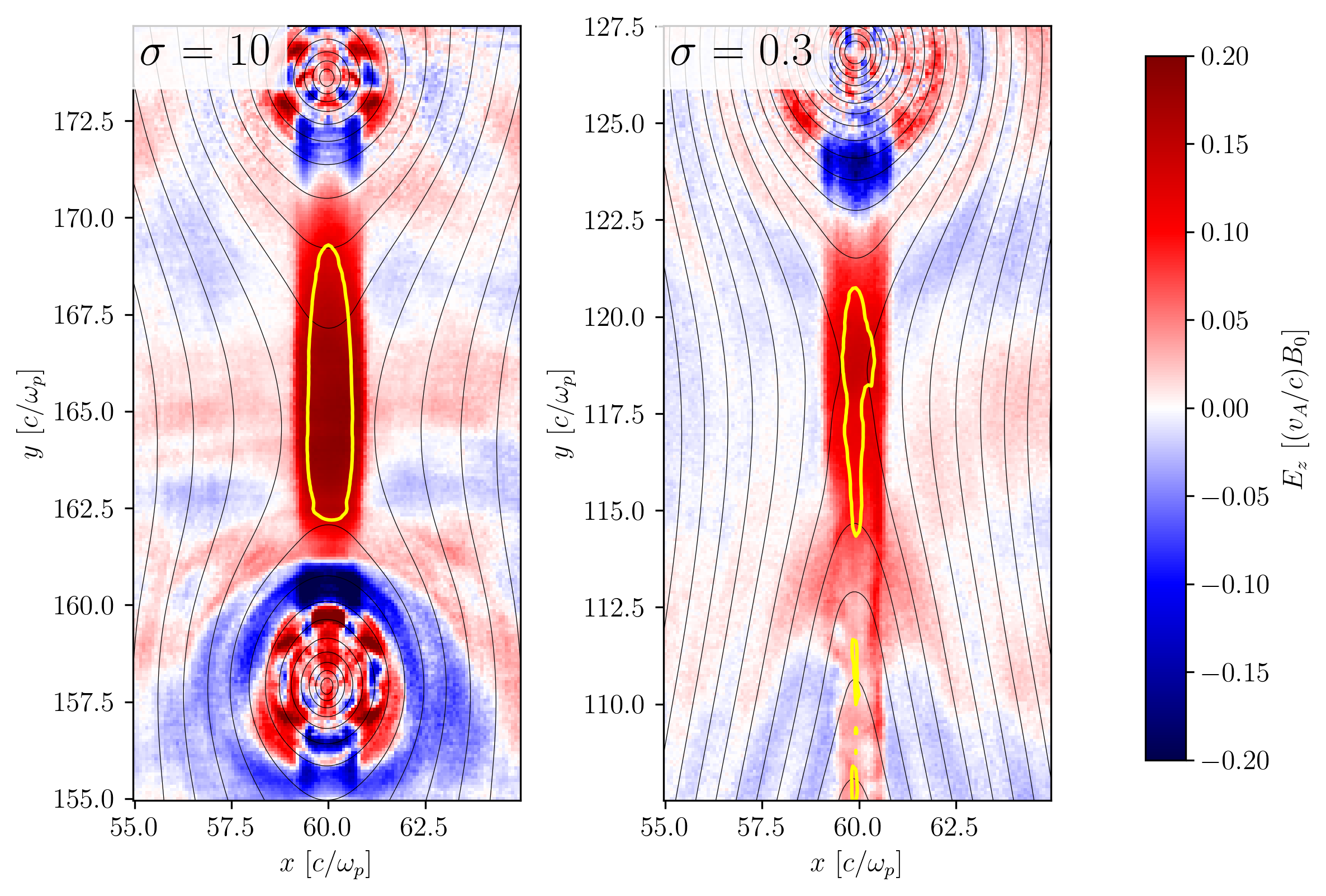}
    \caption{X-points from Runs I and II. The non-ideal electric field $\mathbf{E}_{ni} \equiv \mathbf{E} + \mathbf{v} \times \mathbf{B} / c$ in the out-of-plane direction is plotted. The $E = (v_A / c) B$ contours, giving the boundaries of the diffusion region for the purposes of computing the exit spectra, are shown in yellow.}
    \label{fig:xpts}
\end{figure}

Let us first consider the process by which the initial power law is formed at the X-point for $\sigma = 0.3$ and, specifically, how it differs from that of the $\sigma = 10$ case. We follow the procedure of \citet{bessho2012} (hereafter BB12) who derived the exit energy spectrum of particles leaving the X-point in the fully-relativistic case. Particles which enter at the center of the diffusion region take the longest to be advected out by the $B_x$ field and thus gain the most energy, while those entering near the edges gain the least. To ensure a fair comparison, we compare the primary X-point in the $\sigma = 0.3$ case (here the only X-point) with the primary X-point in the $\sigma = 10$ case. In BB12 the authors define the edge of the diffusion region as the $E = B$ contour, such that the diffusion region is where the electric field dominates over the magnetic field. This is unsuitable in the $\sigma = 0.3$ case since there the $E > B$ zone only covers a small sliver of the diffusion region. We use the modified boundary $E = (v_A / c) B$ since this will define the region in which the electric field term in the Lorentz force law will be greater than that of the magnetic field for a particle leaving the diffusion region at the Alfvén speed (Figure \ref{fig:xpts}). \remove{Since for $\sigma = 10$ the Alfvén speed is highly relativistic, $v_A = 0.95 c$, this leads to nearly the same condition for the high-magnetization case but correctly allows for a larger diffusion region for low magnetization (Figure \ref{fig:xpts}).} As in BB12, we consider a particle as having exited the diffusion region if it passes the maximum or minimum $y$-value of the $E = (v_A / c) B$ contour at that point in time. The exit spectrum is thus the time-integrated distribution of exit energies.

\begin{figure}
    \centering
    \includegraphics[width=0.75\linewidth]{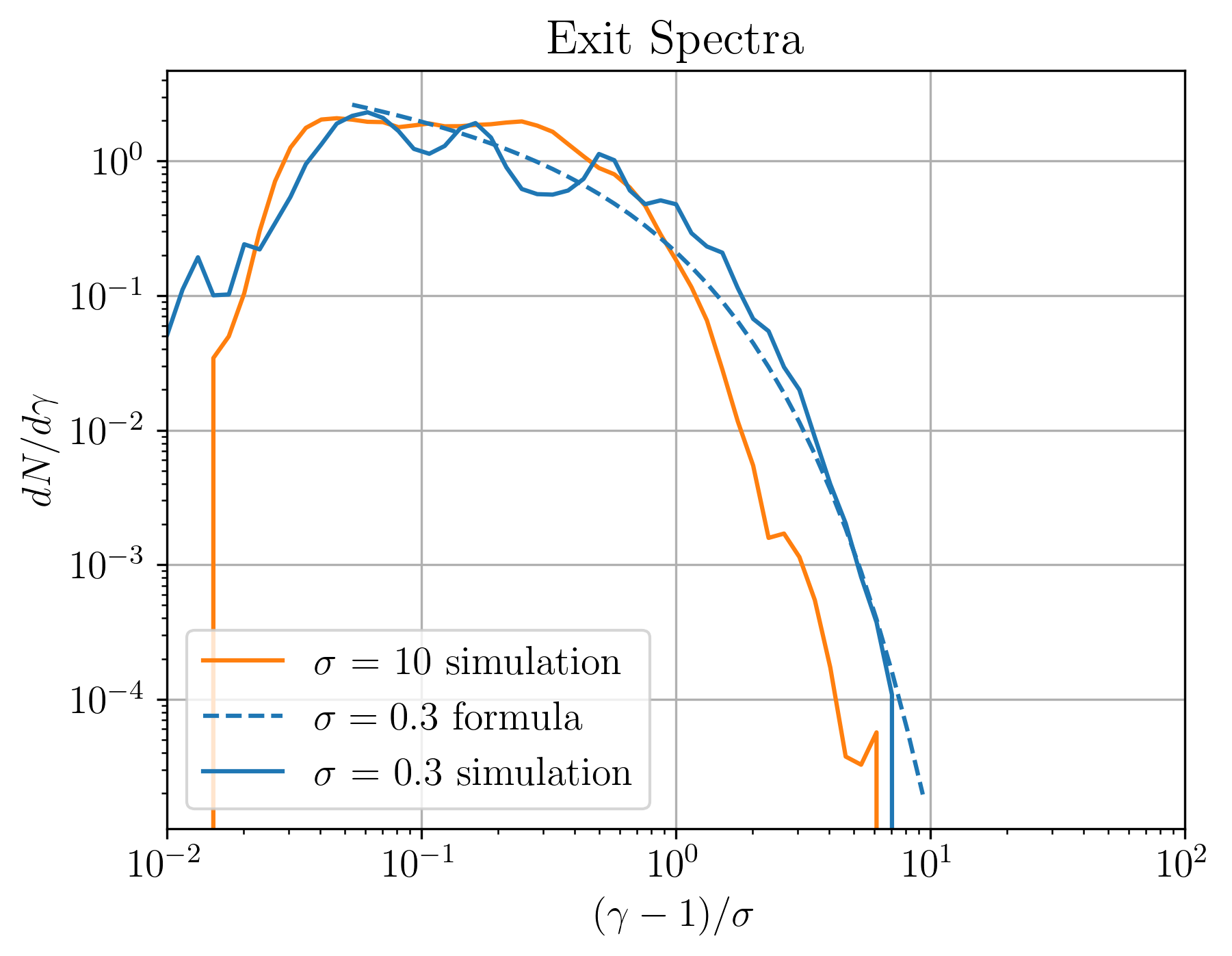}
    \caption{Normalized kinetic energy values for tracked simulation electrons upon their departure from the primary diffusion region, summed over time. The dotted line shows the analytical expression for the nonrelativistic exit spectrum (Equation (\ref{eq:nrspec})). The humps seen in the curve for $\sigma = 0.3$ are caused by a Hall-like $E_x$ component noted by \citet{swisdak2008} which results in step-wise acceleration and bunches up the electron distribution; see Appendix \ref{app:hall} for details. From left to right, the peaks correspond to particles which complete one, two, and three Speiser oscillations before leaving the diffusion region.}
    \label{fig:exit_spectra}
\end{figure}

The exit spectra are shown in Figure~\ref{fig:exit_spectra}. As one can see, in both cases we observe hard spectra---subrelativistic X-points accelerate particles as effectively as relativistic ones do. In fact, particles gain slightly more energy in the $\sigma = 0.3$ case relative to the magnetic free energy available. This supports our presumption that there is nothing inherently special about $E > B$ regions for the nonrelativistic case, since particles can still effectively sample the non-ideal field while $B > E$, since $v / c$ is correspondingly lower. 

If we, like BB12, neglect the Speiser motion in the $x$-direction and repeat their procedure to compute an analytical form for the exit spectrum in the nonrelativistic limit, we arrive at the following result which resembles that of \citet{bulanov76}:

\begin{equation}
    \frac{dN}{dW} \propto (W^*)^{- 1 / 4} \exp{\left(- \frac{2}{3} \varepsilon^{1/2} (2 W^*)^{3 / 4}\right)}. \label{eq:nrspec}
\end{equation} Here $\varepsilon \equiv \sqrt{q E_z^2 / b m c^2}$ is a dimensionless parameter quantifying the relative influence of $E_z$ on the particle trajectory compared to $B_y$,  $W^* \equiv (q E_z^2 / b)^{-1} W$ is the normalized kinetic energy, and $b \equiv \partial B_x / \partial y$, assumed to be constant. Details of the derivation can be found in Appendix \ref{app:deriv}. This fits the data neatly (Figure~\ref{fig:exit_spectra}) and may further explain \remove{the} experimental observations of exponential cutoffs in electron energy spectra \citep{fox2010, na2023}.

\begin{figure}
    \centering
    \includegraphics[width=1\linewidth]{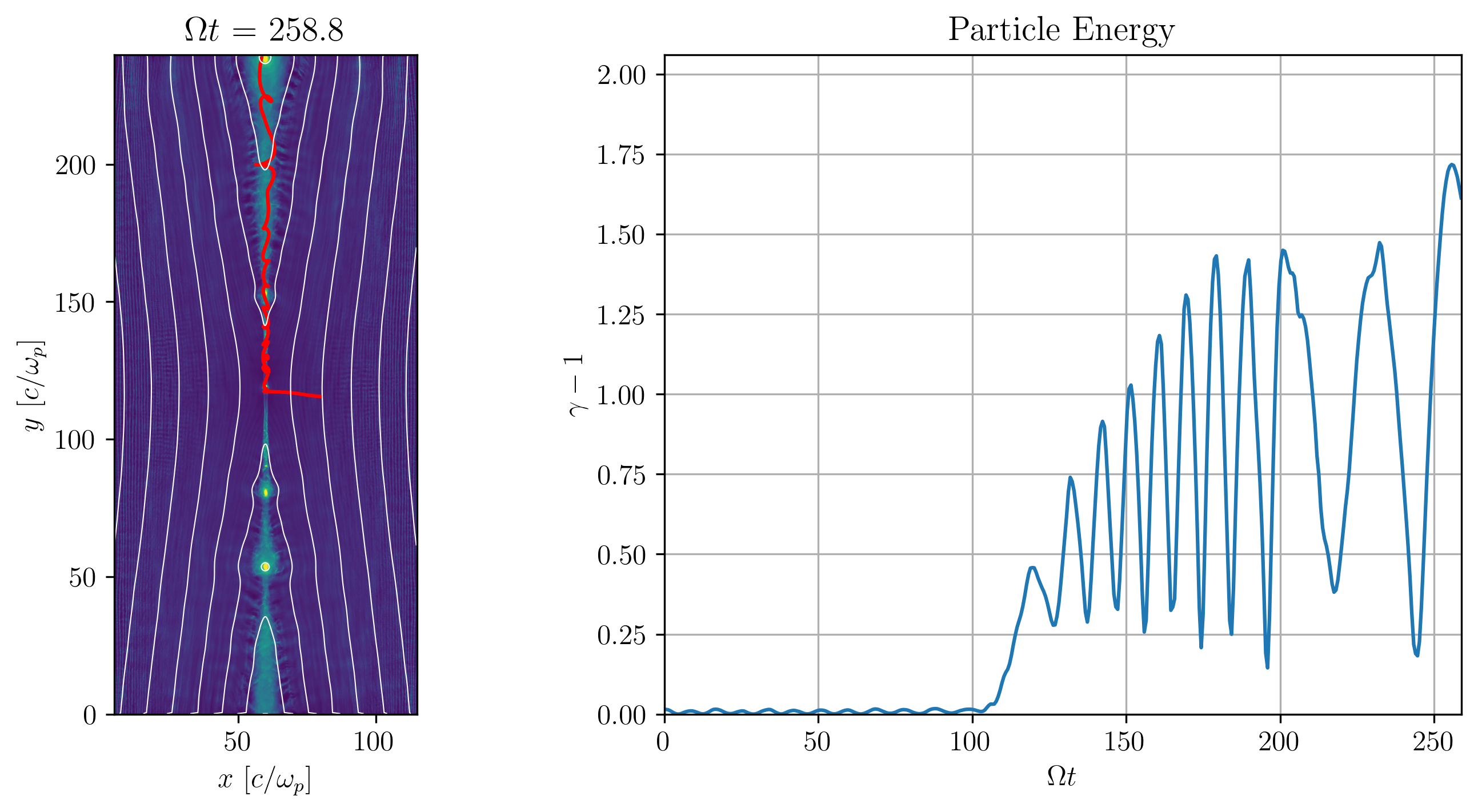}
    \caption{Trajectory of an electron over the density profile for the last timestep plotted for Run I ($\sigma = 0.3$). $\Omega$ is the upstream electron gyrofrequency. The particle first interacts with the current sheet at the X-point \add{($105 \lesssim  \Omega t \lesssim 120$)} and is thereafter confined within a plasmoid \add{($\Omega t \gtrsim 120$)}. \add{As the plasmoid and trapped electron move together into the outflow, the electron energy initially imparted by} direct $E$-field energization is magnified several times by Fermi/betatron heating. The oscillation in energy reflects the gyration up and down the $E_z$ potential.}
    \label{fig:twostep_accel}
\end{figure}

\par 
Once this power law is established, it is enhanced by acceleration within \add{a plasmoid}. Consider the particle shown in Figure~\ref{fig:twostep_accel}, a typical case. This electron gains its first substantial amount of energy upon interaction with the X-point. However, most of its ultimate energy gain comes from being confined within a \add{single} plasmoid as it compresses and its magnetic fields intensify \citep{guo2019, hakobyan2021}. This energization can persist so long as the plasmoid continues to accrete more flux and compress adiabatically, heating the particles within.  Once the plasmoid \add{is} completely immersed within the exhaust, \add{it is} deprived of fresh magnetic flux, starving this acceleration mechanism of additional free energy. As a result, the vast majority of plasmoid heating occurs while the plasmoid is still close to the X-point: past $\gtrsim 1000 \; \skindepth$ skin depths, the plasmoids no longer grow and the particles within the plasmoid subsequently no longer gain energy. Thus, unlike in relativistic reconnection, this acceleration pathway has a fixed window of opportunity and cannot generate arbitrarily high-energy particles provided a large enough domain. 

\subsection{Exhaust Spectrum} \label{sec:sspec}

In the large-domain limit only a vanishingly small fraction of particles interact with the X-point at all. The vast majority are simply heated upon interaction with the exhaust. The basic mechanism is pickup acceleration as described in \citet{drake2009}: when a particle with negligible energy encounters a magnetohydrodynamic flow, the $E$ and $B$ fields accelerate the particle up to speed with the flow. \add{The resultant $E \times B$ cycloid motion imparts a total mean kinetic energy of $m (c E / B)^2 = m v_A^2$, equally partitioned between the gyromotion and the guiding-center motion}. This acceleration process is responsible for the spectral shoulder in Figure~\ref{fig:spectra}\add{, with the mean pickup energy indicated by the dotted line}. Once within the flow, these accelerated particles are re-magnetized and undergo mirror motion along the curved field lines, bouncing between opposite boundaries of the exhaust. \remove{This behavior contrasts with recent results by \citet{walia2022} concerning ions in electron-ion exhausts: there, due to the mass ratio, unmagnetized ions may gain energy reflecting off of shock potentials on either side of the exhaust. Such potentials are absent in pair plasma.}

The net result of this particle energization is a thermal bulk flow proceeding at the Alfv\'en speed, with the energy distribution taking Maxwellian form in the comoving frame \add{with $\theta \equiv T / mc^2 = 0.06$}. Since the vast majority of the total dissipation occurs in these exhausts and not in X-points/plasmoids, we thus see that while high-$\sigma$ reconnection routinely generates strong power laws in energy, low-$\sigma$ reconnection merely heats the inflows (though a small, energetic contribution within the plasmoids is present). This is consistent with recent work emphasizing the importance of injection at X-points \citep{sironi2022,totorica2023,gupta2025}: while the particles which happen to be injected there do reach nonthermal energies, the overwhelming majority which do not will simply be part of the thermal bulk. \remove{Since injected particles are a minuscule part of the total for single X-point reconnection, heating dominates over nonthermal particle acceleration.}

\section{Discussion and Conclusions} \label{sec:conclusion}

We find that when the magnetization $\sigma$ in pair reconnection is reduced below 1, the diminished plasmoid formation rate replaces the self-similar plasmoid chain seen in relativistic reconnection with a laminar Petschek configuration. Below a critical threshold, the plasmoid production rate is too slow to form large plasmoids via mergers, allowing a laminar exhaust structure to form. This exhaust structure increasingly dominates over the plasmoids as the system size increases, although the reconnection rate remains fast. We identify a dimensionless parameter $\pparam$ as a function of the current sheet width, plasmoid production rate, and Alfv\'en speed which quantifies whether a reconnection geometry will tend towards a plasmoid-dominated multiple X-point configuration or an exhaust-dominated single X-point configuration and determine the transition range. 

We then apply these results to explain the difference in particle spectra. For $\sigma < 1$ the strong particle acceleration and power law formation associated with relativistic reconnection is replaced with thermal heating of particles through pickup acceleration. We show via simulations and justify analytically that the X-points in the low-$\sigma$ regime are just as effective at accelerating particles as in the high-$\sigma$ regime: the reason this ultimately matters little is that few particles have the chance to interact with them. A tabular summary of the basic logic and conclusions is given in Table \ref{tab:summary}.

\begin{table}
    \centering
    \begin{tabular}{|l||l|l|}
        \hline
         \textbf{Property} & $\sigma < 1$  & $\sigma \gg 1$ \\
         \hline
         plasmoid parameter $\pparam$ & $\pparam < \pparam_c$ & $\pparam > \pparam_c$ \\
         \hline
         plasmoid mergers & rare & common \\
         \hline
         global geometry & single X-point & multiple X-point\\
         \hline
         plasmoids & insignificant & dominant \\
         \hline
         plasmoid growth & limited & boundless \\
         \hline
         exhausts & dominant & insignificant \\
         \hline
         X-point power law & yes & yes \\
         \hline
         energy spectrum & Maxwellian & hard power law \\
         \hline
    \end{tabular}
    \caption{Summary of our conclusions contrasting the different properties observed in low-$\sigma$ reconnection ($\sigma = 0.3$) and high-$\sigma$ reconnection ($\sigma = 10$). The causal factor responsible for these changes is the increase in the normalized plasmoid production rate $\pparam \equiv \pflux w_{cs} / v_A$ from low to high magnetization.}
    \label{tab:summary}
\end{table}

We have three main findings: 1) in kinetic plasmas, the transition from multiple X-point to single X-point reconnection depends on the rate of plasmoid production in diffusion regions, 2) strong plasmoid activity is necessary for robust power law formation, and 3) stable, fast Petschek reconnection can arise in fully kinetic systems.

Firstly, whether a reconnecting system proceeds as a plasmoid chain or as a Petschek configuration depends on the tearing rate. Either tearing produces sufficiently few plasmoids ($\pparam < \pparam_c$) that they uneventfully advect away from the X-point, resulting in a single X-line geometry, or it produces sufficiently many ($\pparam > \pparam_c$) that they undergo repeated mergers and thereby form a well-developed plasmoid chain. Hence, to predict whether a given system will, in the macroscopic limit, tend towards plasmoid-dominated or exhaust-dominated reconnection, one need only analyze the dynamics near a single X-point and compute $\pparam$. We expect the conclusions reached here for pair plasma to also apply in the electron-ion regime: while mass ratio effects complicate the tearing physics, the same basic macroscale dichotomy remains. This framework can thus explain when one would expect to see an exhaust jet as opposed to a plasmoid chain in the heliosphere or in other environments.

Secondly, the transition between the plasmoid- and exhaust-dominated reconnection is key for predicting particle acceleration. Although debate remains regarding the relative importance of direct $E$-field acceleration as opposed to Fermi mechanisms \citep[e.g.,][]{guo2025, french2025}, our results demonstrate the necessity of plasmoid chains for power law generation. By allowing for stochastic acceleration at multiple X-points and within merging and compressing plasmoids, particles in relativistic reconnection can rapidly form a power-law distribution, avenues that are highly suppressed in an exhaust geometry. This has important effect on understanding whether nonrelativistic electron-ion reconnection is capable of generating strong power laws like those seen in simulations of high-$\sigma$ pair plasmas \citep[e.g.,][]{li2021}. Since the injection physics is insensitive to magnetization (Section \ref{sec:xspec}), we suggest that the answer will depend on whether the secondary tearing rate is high enough to generate frequent plasmoid mergers. Whether this would be observed in simulations depends on whether the box sizes are large enough to capture such dynamics. 

Lastly, our findings provide support for Petschek's reconnection model. While there has been debate over whether Petschek reconnection can be realized from first-principles simulations (i.e., without ad hoc anomalous resistivity), here we demonstrate a simple such case in which clear, stable laminar Petschek outflows form. While previous work suggested similar conclusions \citep{swisdak2008, liu2012, innocenti2015}, ours extends these results in two main ways. Because we use outflow boundary conditions, we can wait for our simulations to reach a steady state, neither a transient nor sensitive to the initial conditions. Also, by simulating much larger domains (thousands of $c/ \omega_p$), we confirm that plasmoid activity becomes less, not more, impactful as the box size is increased, meaning that these results can be extended to macroscopic system sizes. This bolsters the argument that the anomalous resistivity invoked in MHD simulations and theories of reconnection can indeed result from a fully self-consistent kinetic treatment. Ultimately, our results provide a physical framework for explaining the slow-mode shocks observed in the heliosphere as consequences of Petschek reconnection \citep[e.g.,][]{ walia2024}.

Regarding the localization of the diffusion region, our simulations agree with the suggestion of \citet{daughton2007} that it is the tearing instability which prevents the elongation of the diffusion region by fragmenting the thin current sheet at the X-point should it grow too large. This localization then enables fast Petschek-type reconnection. Hence, although plasmoids are too few in number to generate a plasmoid chain in subrelativistic reconnection, they do play an important role in regulating the structure of the diffusion region. 

We leave to further research the extension of these results to electron-ion plasmas and the important question of the role played by three-dimensional effects. Ultimately, the question of whether 3D reconnection proceeds as a dynamic multiple X-point phase or as a predominantly laminar single X-point configuration remains, although flux-rope breakdown \citep[e.g.,][]{zhang2021} \add{and the drift-kink instability \citep[e.g.,][]{bacchini2025}} complicate the simple 2D picture.

\begin{acknowledgments}
The authors thank H.~Ji, Y.~H.~Liu, N.~Walia, and S.~Zenitani for useful discussions. We thank R. Ewart for help with visualization of field lines. We are further grateful for H.~Hakobyan's assistance with TRISTAN V2. \add{We thank the anonymous reviewer for helpful suggestions which clarified the manuscript.} A.R.~acknowledges support from NASA (80HQTR21T0105), and A.S.~acknowledges the support of grants from NSF (PHY-2206607) and the Simons Foundation (MP-SCMPS-00001470). \add{This research was supported in part by grant NSF PHY-2309135 to the Kavli Institute for Theoretical Physics (KITP).}
\end{acknowledgments}

\appendix

\color{\addcolor}

\section{Outflow Boundary Conditions} \label{app:bcs}

The outflow boundary conditions smoothly transition the electric and magnetic field components to target values near the top and bottom of the domain according to

\begin{align}
    A &= e^{-\lambda} A_\text{boundary} + (1 - e^{-\lambda}) A_{\text{target}}, \\
    \lambda &\equiv \kappa \left| (y - y_\text{boundary}) / \dabs \right|^3,
\end{align}
where $A$ is a component of either the electric or magnetic field, $\dabs$ is the absorption length, and $\kappa$ is a strength parameter. For a distance of $\dabs$ cells from the domain edges in the $y$ direction the fields are set to the target value, and the transition occurs over another $\dabs$ from that fixed region. For the electric field the target is $(0, 0, 0.1 B_0)$, and for the magnetic field the target is the initial value thereof. In our simulations, $\kappa = 10$ and $\dabs = 10$.

\color{black}

\section{Derivation of Nonrelativistic Exit Spectrum} \label{app:deriv}

We consider a nonrelativistic electron released from rest subject to $B_x =  b y$ and uniform $E_z$. The equations of motion are thus

\begin{align}
	m \ddot{y} &= q \frac{\dot{z}}{c} b y, \\
	m \ddot{z} &= - q E_z - \frac{q}{c} \dot{y} b y.
\end{align} If we define $\Delta \equiv E_z / b$, $\tau_E \equiv\sqrt{m \Delta / q E_z}$, $\Omega \equiv q b \Delta / m c$, and normalize $t$ by $\tau_E$ and $y, z$ by $\Delta$, we obtain

\begin{align}
	\ddot{y} &= \varepsilon y \dot{z}, \label{eq:futureairy} \\
	\ddot{z} &= -1 - \varepsilon y \dot{y},
\end{align} where $\varepsilon \equiv \tau_E \Omega$. For $\varepsilon \ll 1$, we can take $z = -t^2 / 2$ and use this to solve the resulting Airy equation Equation (\ref{eq:futureairy})

\begin{align}
	\ddot{y} = \varepsilon y t.
\end{align} Considering an electron starting from rest, $\dot{y}(0) = 0$, we find

\begin{align}
	y = y_0 A \left(-\frac{\Ai{\zeta}}{\Aip{0}} + \frac{\Bi{\zeta}}{\Bip{0}}\right),
\end{align} with $\zeta \equiv \varepsilon^{1/3} t$ and where $A$ must be

\begin{align}
	A = \left(-\frac{\Ai{0}}{\Aip{0}} + \frac{\Bi{0}}{\Bip{0}}\right)^{-1}.
\end{align} We wish to find the time at which $y = y_f$, where $y_f$ is the edge of the diffusion region. If we presume that this particle enters near the center of the diffusion region and thus remains therein for a significant amount of time, we can use the large-$t$ limit to drop the Airy function first kind and approximate the Airy function of the second kind as
\begin{align}
	\Bi{\zeta} \approx \frac{1}{\sqrt{\pi} \zeta^{1 / 4}} \exp{\left(\frac{2}{3} \zeta^{3 / 2}\right)}.
\end{align} These simplifications yield the $y = y_f$ condition for $t$ as
\begin{align}
	y_0 = y_f (2 \sqrt{\pi} \Bi{0}) \zeta^{1/4} \exp\left({-\frac{2}{3} \zeta^{3 / 2}}\right).\label{eq:yinityfinal}
\end{align} The final spread in energy will be due to particles entering the diffusion region at different $y_0$ as
\begin{align}
	\frac{dN}{dW_f} = \frac{dN}{dy_0} \frac{dy_0}{dW_f},
\end{align} where $W_f = t_f^2 / 2$ is the dimensionless kinetic energy when $y = y_f$. Substituting Equation (\ref{eq:yinityfinal}) and assuming $dN / dy_0$ is constant (i.e., a spatially uniform inflow density) yields our energy spectrum:
\begin{align}
    \frac{dN}{dW_f} \propto W^{-1/4} \exp{\left(-\frac{2}{3} \varepsilon^{1 / 2} (2 W)^{3 / 4}\right)}.
\end{align} This is similar to that found by \citet{bessho2012} in the relativistic limit, though there the $W$ dependence in the exponential is $W^{1/2}$ instead of $W^{3 / 4}$.

\section{Hall-like Fields in the Diffusion Region} \label{app:hall}

\begin{figure}
    \centering
    \includegraphics[width=0.25\linewidth]{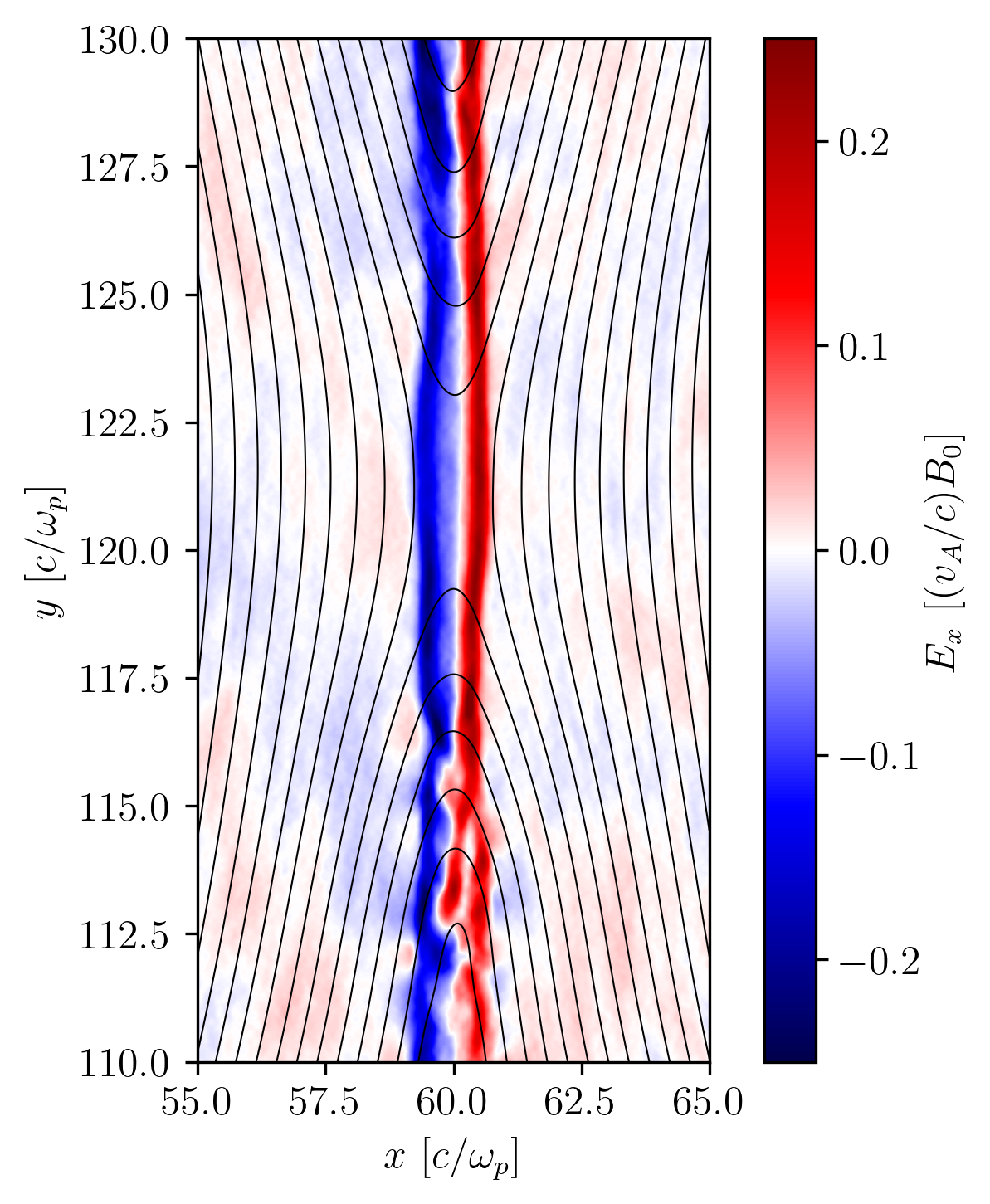}
    \caption{Hall-like electric fields near the diffusion region, with magnetic field lines overlaid in black.}
    \label{fig:ess}
\end{figure}

At the center of the $\sigma = 0.3$ reconnection domain lies a tearing-unstable current sheet at the X-point. Given the unity mass ratio, charge separation and the establishment of Hall fields would not be expected. However, strong electrostatic fields comparable in strength to the reconnection electric field are a robust feature of $\sigma = 0.3$ systems, appearing in all such simulations conducted (Figure~\ref{fig:ess}). This field structure was noted earlier by \citet{swisdak2008} in similar simulations. Like Hall fields, these electric fields are directed along the $x$-direction. Due to the unity mass ratio, either electric field polarity is possible: for definiteness, we take the ion-like species to be electrons and the electron-like species to be positrons, as develops in Run I. This causes electrons to undergo step-like acceleration as they gain and lose energy to this electric potential over the course of their Speiser orbits, resulting in the discrete bunching of particle energies as seen in Figure~\ref{fig:exit_spectra}. It also results (in this polarity) in the electrons possessing more energy than the positrons since they are attracted by the potential. The shapes of the resultant electron and positron energy distributions are nearly identical, with the electron energies simply shifted to the right in the final spectrum. 
This effect is entirely absent in the $\sigma = 10$ case. 

\bibliography{references}{}
\bibliographystyle{aasjournalv7}

\end{document}